\documentclass[sigconf]{acmart}

\renewcommand\footnotetextcopyrightpermission[1]{}
\AtBeginDocument{%
  \providecommand\BibTeX{{%
    \normalfont B\kern-0.5em{\scshape i\kern-0.25em b}\kern-0.8em\TeX}}}

\usepackage{algorithm}
\usepackage{algorithmic}
\usepackage{multirow}
\usepackage{xspace}
\usepackage{enumitem}
\usepackage{bbding}
\usepackage{stfloats}
\usepackage{makecell}
\usepackage{minted}

\usepackage{tabularx}

\newcommand{\ie}{{\em i.e.},\xspace}
\newcommand{\eg}{{\em e.g.},\xspace}

\newcommand{\miohint}{\textsc{Mio}\textsc{Hint}\xspace}

\begin{document}

\title[LLM-Assisted Request Mutation for Whitebox REST API Testing]{MioHint: LLM-Assisted Request Mutation for Whitebox REST API Testing}

\author{Jia Li}
\affiliation{%
  \institution{The Chinese University of Hong Kong}
  \city{Hong Kong}
  \country{China}
  }
\email{linsayli@link.cuhk.edu.hk}

\author{Jiacheng Shen}
\affiliation{%
  \institution{Duke Kunshan University}
  \city{Kunshan}
  \country{China}
  }
\email{jc.shen@dukekunshan.edu.cn}

\author{Yuxin Su}
\affiliation{%
  \institution{Sun Yat-sen University}
  \city{Zhuhai}
  \country{China}
  }
\email{suyx35@mail.sysu.edu.cn}

\author{Michael R. Lyu}
\affiliation{%
  \institution{The Chinese University of Hong Kong}
  \city{Hong Kong}
  \country{China}
  }
\email{lyu@cse.cuhk.edu.hk}


\begin{abstract}
Cloud applications heavily rely on APIs to communicate with each other and exchange data.
To ensure the reliability of cloud applications, cloud providers widely adopt API testing techniques.
Unfortunately, existing API testing approaches are insufficient to reach strict conditions, a problem known as fitness plateaus, due to the lack of a gradient provided by coverage metrics.
To address this issue, we propose \miohint, a novel white-box API testing approach that leverages the code comprehension capabilities of Large Language Model (LLM) to boost API testing.
The key challenge of LLM-based API testing lies in system-level testing, which emphasizes the dependencies between requests and targets across functions and files, thereby making the entire codebase the object of analysis. However, feeding the entire codebase to an LLM is impractical due to its limited context length and short memory.
\miohint addresses this challenge by synergizing static analysis with LLMs.
We retrieve relevant code with data-dependency analysis at the statement level, including def-use analysis for variables used in the target and function expansion for subfunctions called by the target.

To evaluate the effectiveness of our method, we conducted experiments across 16 real-world REST API services. 
The findings reveal that \textsc{MioHint} achieves an average increase of 4.95\% absolute in line coverage compared to the baseline, \textsc{EvoMaster}, alongside a remarkable factor of 67$\times$ improvement in mutation accuracy. 
Furthermore, our method successfully covers over 57\% of hard-to-cover targets, while in the baseline, the coverage is less than 10\%.
\end{abstract}


\maketitle

\section{Introduction}\label{sec:introduction}
The proliferation of cloud-based applications and services has led to an exponential increase in the reliance on APIs for communication and data exchange, particularly in RESTful architectures. Consequently, ensuring the reliability, security, and performance of RESTful APIs through automated testing has become a critical aspect of software development and deployment. 

Despite its importance, the majority of automated testing efforts are still centered around black-box testing~\cite{10.1109/ICSE.2019.00083,liu2022morestmodelbasedrestfulapi,9159077,golmohammadi2022testingrestfulapissurvey}, which typically achieves low coverage~\cite{zhang2022openproblemsfuzzingrestful}. 
Black-box API testing treats the API as a closed system, focusing solely on the inputs and outputs without any knowledge of the internal workings of the API. 
While this approach is valuable for validating the external behavior of the API, it often falls short in dealing with corner cases and deep system states.

On the other hand, white-box API testing leverages runtime information (\eg coverage) of the APIs by instrumentation of source code. With this information, white-box testing defines heuristics and applies search-based techniques to fuzz REST APIs. Specifically, it generates new test cases by randomly mutating existing test cases, and then evaluates the new test cases with its code coverage. This approach achieves better results in code coverage due to the coverage-guided search process~\cite{zhang2022openproblemsfuzzingrestful}.
However, white-box approaches fall short when encountering fitness plateaus~\cite{arcuri2021enhancing}, a widely recognized problem when coverage provides no gradient to the search, \eg a strict equality check. In this situation, random mutations become inefficient at improving coverage.

To address these limitations, there is a growing need for more sophisticated white-box testing approaches that leverage a deeper understanding of the codebase. 
One such approach is symbolic execution, which tracks program inputs as symbolic variables and maintains symbolic expressions across statements. This technique allows for the construction of constraints on the input related to the target, which can then be solved using a constraint solver. After solving, the program can directly reach the target with the solution.

However, symbolic execution suffers from model boundaries (\eg external libraries), path explosion, and imprecise abstraction. 
As a result, it struggles to scale effectively and is impractical in complex software systems.
Given these challenges, there is a pressing need for innovative solutions that apply lightweight code analysis techniques capable of managing the demands of large codebases without compromising accuracy.

Large language models (LLMs) present a novel opportunity in the realm of API testing. These models possess the capability to comprehend the semantics of static code snippets and execute a variety of code-related tasks~\cite{liu2024your,10.1145/3650212.3680399}. Unlike heavyweight program analysis techniques such as symbolic execution, which require modeling the program states from the beginning and maintaining all related state information—resulting in a significant increase in state space and analysis overhead as the size of the codebase grows, LLMs can perform localized analysis on code snippets extracted from a large codebase. This approach ensures that the analysis overhead for each LLM query remains nearly constant, regardless of the size of the codebase.

However, utilizing LLMs for system-level API testing presents significant challenges.
The difficulty arises from the necessity to consider the data-dependency relationships between inputs (\eg HTTP requests) and the target. 
These relationships span the entire code repository, thereby making the entire repository the object of analysis. 
Given the limited context length and short memory of LLM's reasoning, it is impractical to incorporate the entire repository into a single prompt.

Thus, there is a need for inter-file and inter-procedural analysis techniques to extract code relevant to the target to produce a global context. Retrieval-based approaches~\cite{lu2022reaccretrievalaugmentedcodecompletion,zhang2023repocoderrepositorylevelcodecompletion} that identify related references based on identifier names and semantic similarity overlook the intrinsic structures of programming languages, such as function call chain and data dependency graph, which can lead to low accuracy.
State-of-the-art code extraction approaches enhanced by static analysis~\cite{chen2024chatunitestframeworkllmbasedtest,phan2024repohyper,ding2023cocomiccodecompletionjointly,ma2024understand} aim to address these limitations. However, these approaches operate at the granularity of methods or classes, which can introduce significant amounts of redundant information and further reduce accuracy.

To address the common issue of fitness plateaus in search algorithms, we enhance mutation accuracy with a lightweight analysis provided by the LLMs. Ideally, high-accuracy mutations can directly reach hard-to-cover targets, thereby overcoming the entrapment in local optima imposed by fitness plateaus.
Given the vast information landscape presented by system-level testing, specifically the entire repository, we retrieve global context through statement-level data dependency analysis to effectively capture the relevant context while minimizing redundancy.

Our proposed approach, \textsc{MioHint}, initiates LLM-assisted mutation when the search algorithm encounters hard-to-cover targets. 
This integrated approach creates a synergistic system that balances search effectiveness with efficiency. It leverages the fast, broad exploration capabilities of the search algorithm for general progress, while strategically deploying the targeted LLM-assisted mutation as a powerful specialist to overcome difficult coverage targets that would otherwise cause the search to stagnate. This division of labor effectively optimizes the trade-off between search breadth and depth, thereby maximizing overall testing performance.
In the implementation of LLM-assisted mutation, for each target, \textsc{MioHint} begins by performing statement-level value expansion to extract global context from the entire codebase. It then constructs a sophisticated prompt by synthesizing this global context with local function code and existing requests. This prompt strategically employs chain-of-thought reasoning to guide the LLM in methodically locating the field for revision and in-context learning to incorporate feedback from previous failed mutations. Finally, this structured prompt is used to query the LLM for precise mutation guidance.

We integrate \textsc{MioHint} into \textsc{EvoMaster}~\cite{arcuri2021evomaster}, a widely adopted whitebox API testing framework. 
We evaluate it with 16 real-world REST API services of EMB~\cite{emb}, which add up to 314,415 lines of code. 
The experiments demonstrate that compared to the baseline \textsc{EvoMaster}, our approach increases line coverage by an average of 4.92\% and achieves a 67$\times$ increase in mutation accuracy. Besides, our approach successfully covers over 57\% of hard-to-cover targets, while in the baseline, the coverage is less than 10\%.

In summary, we make the following contributions in this paper:
\begin{itemize}
    \item We propose \textsc{MioHint}, which integrates LLM-assisted mutation with an advanced search algorithm. It makes a substantial improvement on the accuracy of mutation with the aid of LLMs.
    \item We propose a statement-level data dependency code extraction approach, that addresses the inaccuracy issue of current method-level approaches.
    \item We conduct a large-scale evaluation of \textsc{MioHint} in variant real-world web services. The results demonstrate that our \miohint significantly increases the accuracy of mutation and then improves the line coverage of programs under test.
    \item We release our open-source implementation of \textsc{MioHint} and the associated data to help replicate the experiments in this paper~\cite{datapublic}.
\end{itemize}

\begin{figure*}[htbp]
    \centering
    \includegraphics[width=0.9\textwidth]{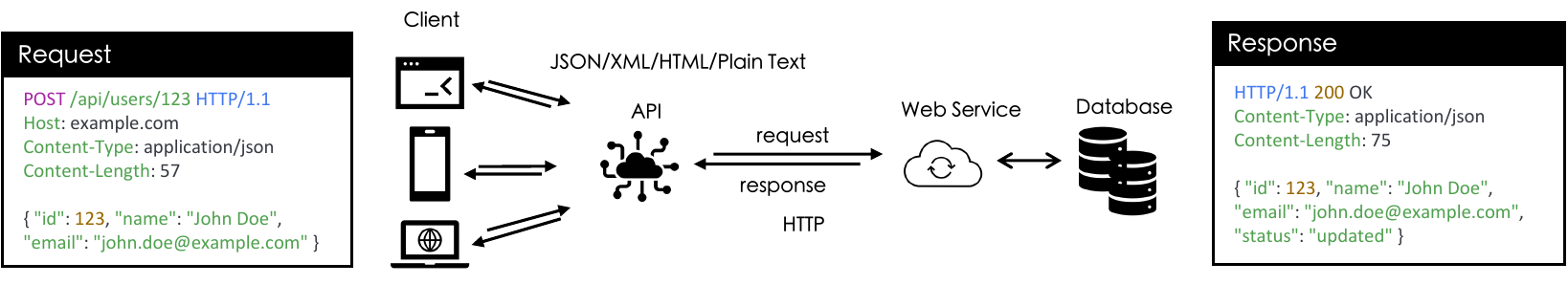}
    \caption{Web Service and API.}
    \label{fig:webservive}
\end{figure*}

\section{Background and Motivation}\label{sec:background}

\subsection{Web Services and API}
Web services and APIs (Application Programming Interfaces) are essential components in modern software development. 
As shown in Figure~\ref{fig:webservive}, APIs provide standardized methods for applications to exchange data and functionality, regardless of the underlying platforms or technologies. 

The REST API (Representational State Transfer Application Programming Interface) is the most widely adopted API specification. 
REST API is based on the HTTP protocol. 
It typically employs standard HTTP methods (GET, POST, PUT, DELETE) to define operations, uses URIs to identify resources, leverages standard HTTP status codes (such as 200, 404, 500) to indicate request outcomes, utilizes JSON or XML for data interchange, and incorporates HTTP headers (like Content-Type and Authorization) to convey metadata and control information. We listed examples of request and response in REST APIs in Figure~\ref{fig:webservive}, to show how applications exchange data by REST APIs.

\begin{figure}
    \centering
    \includegraphics[width=\columnwidth]{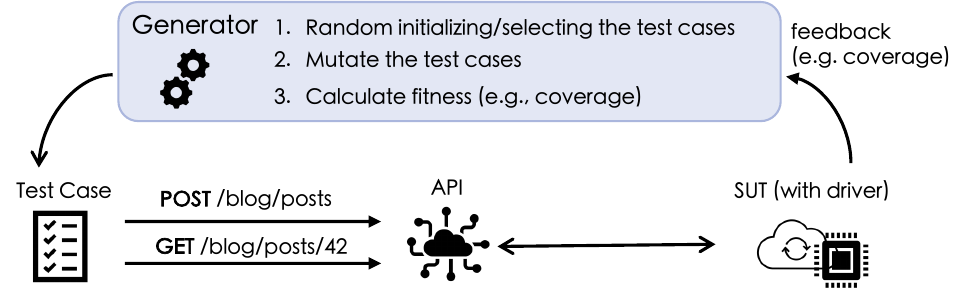}
    \caption{High-level view of Search-based API Testing.}
    \label{fig:sbstapi}
\end{figure}

\subsection{Automated API Testing}
Automated API testing is essential in modern software development to ensure the reliability, performance, and security of the system under test (SUT). 
Existing API testing approaches can be categorized into black-box and white-box approaches. 

\indent \textbf{BlackBox API Testing.} 
Most of the existing API testing tools (\ie 73\%) are black-box~\cite{golmohammadi2022testingrestfulapissurvey}. 
BlackBox API testing conceptualizes the SUTs as an enclosed entity.
It validates the system's behavior by iteratively initiating API requests and verifying the correctness of their responses.
The key challenges for black-box approaches are 1) generating valid inputs and 2) defining appropriate test oracles to evaluate system correctness.
Existing approaches rely on API specifications provided by service providers to generate valid inputs.
They employ machine learning techniques or regular expression matching to interpret the specifications and subsequently generate valid requests~\cite{liu2022morestmodelbasedrestfulapi,10.1109/ICSE.2019.00083}. 
Regarding test oracles, they typically depend on the status codes embedded in API responses to check the system's correctness, \eg 500 refers to an internal server error.

Existing black-box testing approaches often achieve low code coverage because they lack access to the system's implementation.
Consequently, they can only mutate inputs randomly based on the limited information provided in specifications.
This random process frequently fails to cover complex system states, strict constraints, and corner cases.
Furthermore, these approaches rely heavily on accurate and complete specifications from service providers, a requirement that is often not met.
When specifications are inaccurate, these methods tend to generate numerous invalid inputs, which further diminishes code coverage~\cite{zhang2022openproblemsfuzzingrestful,9700203}.
Moreover, the under-specified schema results in all the code related to handling and associated functionalities never being tested if certain HTTP headers and query parameters are not defined in the specification~\cite{arcuri2023advancedwhiteboxheuristicssearchbased}.

\indent \textbf{WhiteBox API Testing.} 
White-box approaches achieve higher code coverage by inspecting the runtime information of SUTs. 
For instance, \textsc{EvoMaster} is the state-of-the-art white-box API fuzzer that adopts code instrumentation to gain the internal visibility into the SUTs~\cite{arcuri2021evomaster}.
Specifically, \textsc{EvoMaster} identifies uncovered code by maintaining an access counter for each line of bytecode.
This counter is incremented each time the corresponding bytecode is executed.
Subsequently, it employs a heuristic-based search algorithm, Many Independent Objects (MIO), which uses the collected bytecode-level coverage information to maximize code coverage.

This category of API testing is known as Search-Based API Testing. Figure~\ref{fig:sbstapi} illustrates a high-level abstraction of this approach. Search-Based API Testing explores the input space by generating new test cases through mutation and utilizes feedback, such as code coverage, to guide the search process effectively.

Specifically, MIO is a variation of the evolutionary algorithm. 
It maintains a pool of high-quality inputs, \ie inputs that have the potential to increase the coverage.
In each iteration, the algorithm randomly selects an input from this pool and applies mutations to it.
The goal of this mutation is to execute different parts of the code.
The code coverage is then measured during its execution.
A mutated input is added to the pool only if it increases the overall code coverage.
The rationale behind this is that such inputs have successfully triggered previously unexecuted code blocks.

However, the code coverage achieved by \textsc{EvoMaster} using MIO is still unsatisfactory.
The fundamental reason is due to the ubiquitous fitness plateaus, \ie mutated test cases are unlikely to increase coverage, and the search process is trapped in local optima~\cite{aleti2017analysing}.

\begin{figure*}
    \centering
    \includegraphics[width=0.9\textwidth]{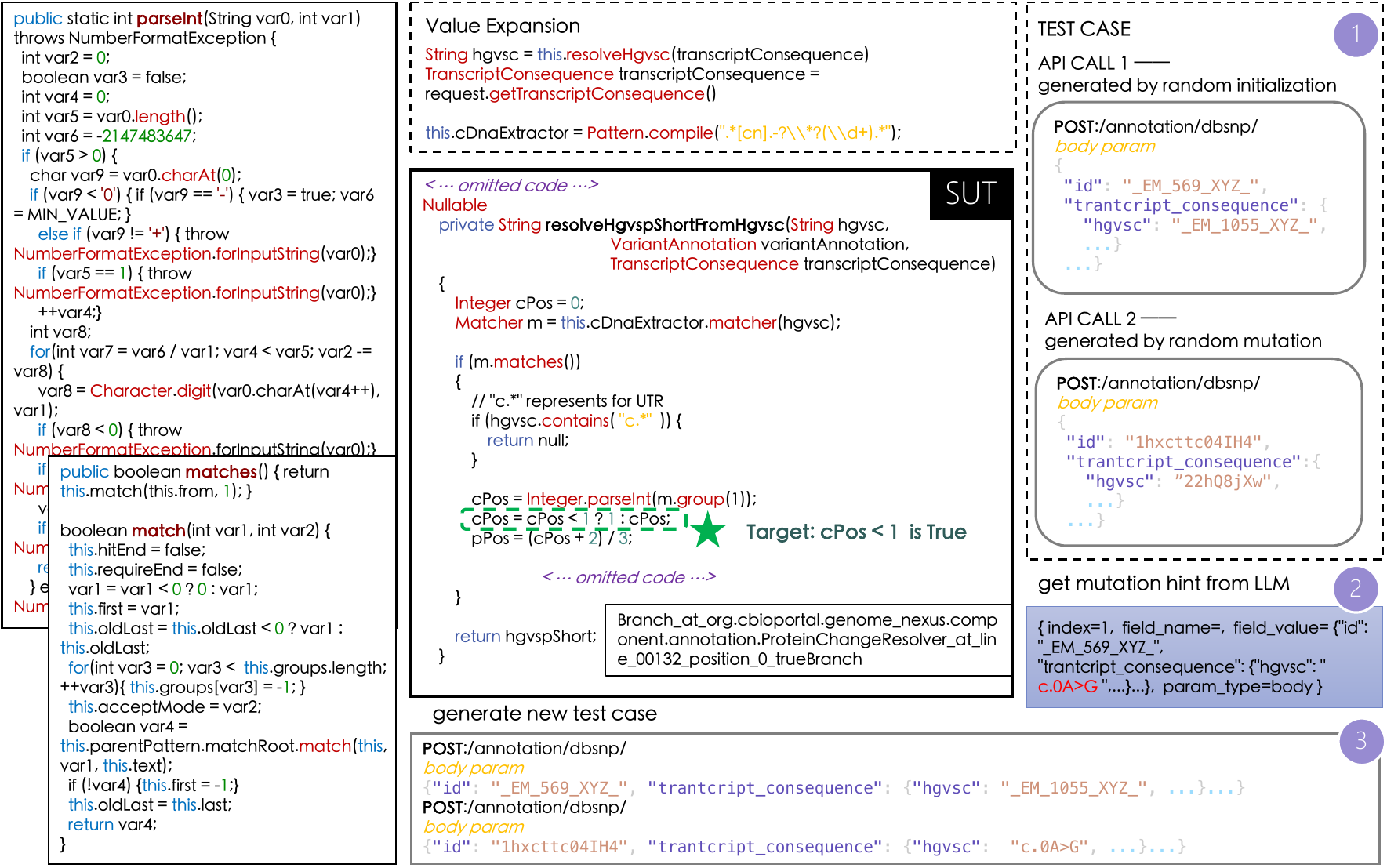}
    \caption{Example run of \textsc{MioHint} on a program under test. When search-based testing encounters fitness plateaus where the mutation is inefficient (1), \textsc{MioHint} queries GPT-4o for a mutation hint (2), and mutates this test case according to the hint (3).}
    \label{fig:motivatingexample}
\end{figure*}

Figure~\ref{fig:motivatingexample} shows an example of the fitness plateau.
The target is to reach condition \texttt{cPos < 1} is \texttt{true}, which is inside function \texttt{resolveHgvspShortFromHgvsc}. 
To cover the target, \texttt{hgvsc} in the request must matches pattern \texttt{.*[cn].-?\textbackslash{}*?(\textbackslash{}\textbackslash{}d+).*} and the number after \texttt{c.} or \texttt{n.} in \texttt{hgvsc} is less than 1. 
Label (1) shows two test cases generated by random initialization and random mutation. 
Random initialization assigns a field with a type string in the form \texttt{\_EM\_\textbackslash{}d+\_XYZ} because of taint analysis in \textsc{EvoMaster}. 
Random mutation in a string involves randomly selecting one or more positions in the string and replacing the characters at those positions with randomly chosen new characters; thus, it generates a sequence of nonsensical characters \texttt{22hQ8jXw}. 
Despite its support for regex-based string generation via testability transformations~\cite{arcuri2021enhancing}, \textsc{EvoMaster} struggles to produce the exact number 0 for the positive number specified in the pattern.
Thus, despite multiple mutations, it remains challenging to produce a mutant that meets the required condition, causing the entire search process to stall at this point. 
Such strict conditions cause numerous spikes in the fitness landscape and continuously hinder the search process.

A straightforward approach is to adopt traditional code analysis techniques, \eg symbolic execution, to overcome the fitness plateaus.
Symbolic execution tracks the semantics in terms of any symbolic input at the bytecode level, formulates constraints, and satisfies them by an SMT solver~\cite{10.5555/1792734.1792766}. 
However, the application of symbolic execution to complex real-world software inevitably encounters several challenges, including model boundaries (e.g., external libraries), path explosion, and imprecise abstraction~\cite{217563,poeplau2020symcc}. 
Specifically, when external libraries are not analyzed, the symbolic execution loses track of symbolic expressions whenever data is passed through a function provided by the library. 
Besides, the number of control-flow paths grows exponentially with an increase in program size, resulting in significant analysis overhead. 
Moreover, any imprecise abstraction of complex data types can lead to incorrect constraints.

The problems with symbolic execution and traditional program analysis techniques call for a lightweight source code analysis approach to attack the problem of fitness plateaus.

\subsection{Opportunity of Code Understanding with LLM}

We propose leveraging the code comprehension capabilities of LLMs to boost the coverage of API testing.
LLMs are auto-regressive models designed to understand and generate natural language by leveraging vast amounts of data (\eg 700 GB). It presents an impressive performance at code understanding and generation~\cite{liu2024your,10.1145/3650212.3680399}.

Using LLMs can effectively address the issue of fitness plateaus without the additional overhead of traditional techniques like symbolic execution.
First, an LLM analyzes fragmented code snippets instead of resolving all data types and dependencies at the bytecode level, thus reducing the analysis overhead. 
Second, LLMs can infer the functionality of external function calls from their names, such as \texttt{matches}, \texttt{parseInt}, and \texttt{group}, instead of performing complex symbolic reasoning on them.

As shown in Figure~\ref{fig:motivatingexample}, we leverage the LLM to generate input for the same example. The LLM is queried with the function code, the target line of code, and an initial input to mutate. Remarkably, the LLM produces the mutation hint shown at label (2), enabling the search to reach the condition \texttt{cPos < 1 == true} in a single query. This is achieved through the LLM’s semantic understanding of the function: it directly generates the string \texttt{c.0A>G} for field \texttt{hgvsc}, which satisfies the target condition.

Finally, a new test case is generated by applying the mutation hint to the selected API call, labeled (3). This new test case is crucial for breaking through the fitness plateau and improving code coverage. This is because it successfully exercises a branch of the code that was previously unreachable by random mutations. 

The LLM-generated hint is effective for two reasons. First, by analyzing data dependencies in the code, the LLM correctly identifies that the input field \texttt{hgvsc} influences the critical variable \texttt{cPos} in the target condition. Subsequently, it determines the precise value \texttt{c.0A>G} required to steer program execution to a previously uncovered branch by satisfying the condition \texttt{cPos < 1}.

Once the high-quality mutant is generated, it is immediately evaluated to determine any increase in fitness. If the fitness improves, the high-quality mutant is saved as a new test case. Subsequently, random mutations can be performed on this new test case to explore other parts of the search space broadly. This approach effectively escapes local optima and yields additional benefits beyond merely covering one target.

\subsection{Challenges of LLM-Based API Testing at Repository Level}
Large language models have been widely adopted for unit test generation due to their remarkable capabilities in code understanding and generation.
To enhance the effectiveness of unit test generation, numerous methods have been proposed~\cite{alagarsamy2023a3testassertionaugmentedautomatedtest,shin2024domainadaptationcodemodelbased,khanfir2023efficientmutationtestingpretrained,chen2024chatunitestframeworkllmbasedtest,yuan2024manualtestsevaluatingimproving,vikram2024largelanguagemodelswrite,plein2023automaticgenerationtestcases,codamosa2023}. 
However, unit testing primarily focuses on isolated functions and small segments of code within a repository and has no concern about other parts of the codebase.
In contrast, our API testing is a system-level testing that emphasizes the dependencies between requests and targets.
These dependencies span across functions and files, thereby constituting a repository-level concern.

Repository-level tasks are challenging as information spans across massive code. State-of-the-art works, including package migration ~\cite{bairi2023codeplanrepositorylevelcodingusing}, issue resolution ~\cite{ma2024understand,xia2024agentless,zhang2024autocoderover}, and code completion ~\cite{phan2024repohyper,ding2023cocomiccodecompletionjointly,shrivastava2023repository}, employ dependency graphs to identify relevant code components through method/class relationships (\eg imports, invocations, inheritance).
However, different from existing repository-level tasks that focus on capturing module interactions, API testing emphasizes value propagation paths between API requests and testing targets.
Specifically, the goal of API testing is to cover the target by modifying the request. Therefore, the key is to determine the target's expression with respect to the request.
Once this expression is established, it provides a constraint representing the target condition. To cover the target, we simply need to satisfy this constraint.
To build the expression with respect to the request, we need to track value transmission at the statement level. 
This analysis works by starting from the variables of the target and iteratively finding their definitions until the variable of the request. 
Previous method-level or class-level dependency graphs, though useful for identifying related modules, introduce extraneous code elements that obscure the critical value transmission chain. 
This over-inclusion occurs because class/method dependencies aggregate all interactions within a component, whereas only a subset of these interactions pertain to the specific request-target relationship.
Considering the accuracy of semantic representation, as well as the limitations on the length of LLM context and cost considerations, we have opted to conduct code extraction at the statement level. 
\section{\miohint}\label{sec:technique}
We propose \miohint to boost the coverage of API testing.
The key idea of \miohint is to utilize the code comprehension capabilities of LLMs to generate mutation hints for hard-to-cover targets.
We further address the challenges of adopting LLMs to repository-level testing with value expansion, a data-dependency static analysis at the statement level.

\begin{figure}
    \centering
    \includegraphics[width=0.8\columnwidth]{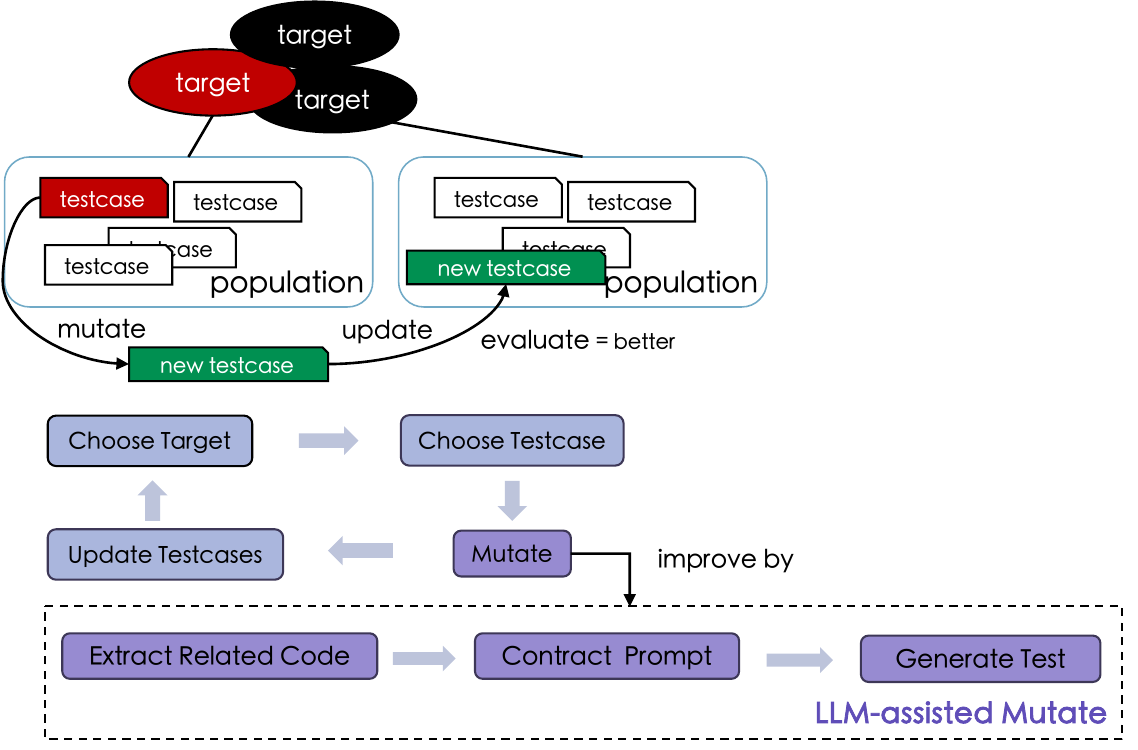}
    \caption{Overview of \textsc{MioHint}’s framework for API test generation.}
    \label{fig:architecture}
\end{figure}

Figure \ref{fig:architecture} shows an overview of how \textsc{MioHint} conducts API testing. 
For each run, we (Step-I) select a target, (Step-II) choose a test case from the target's population, (Step-III) mutate the selected test case, and finally, (Step-IV) update the mutated test case to the population based on its evaluation outcome.

To improve mutation accuracy, we substitute the vanilla mutation with our proposed LLM-assisted mutation. This process involves three main steps, which are related code extraction, prompt construction, and test case generation. Related code extraction retrieves static code related to the target statement from the codebase. Prompt construction combines the information gathered to form a task instruction. Test generation queries LLMs with the constructed prompt, processes the response to get the mutation hint, and applies the mutation hint to the original test case to generate a new one.

In the rest of this section, we first present a high-level overview of \textsc{MioHint}.
We then provide a detailed explanation of the related code extraction, prompt construction, and test generation steps.

\begin{algorithm}
\scriptsize
\caption{\textsc{MioHint}. Parts of the algorithm that are the same as Algorithm MIO are greyed out.}
\label{alg:miohint}
\begin{algorithmic}[1]
\REQUIRE $program$ to test, search time $T$, $model$ to query for test cases, and minimum count of times to query per iteration $M$.
\ENSURE a set of test cases exercising the program that maximizes the coverage.
\STATE \textcolor{gray}{$apis \gets \text{\textsc{GetApis}}(program)$}
\STATE \textcolor{gray}{$testCases \gets \text{\textsc{RandomTestCases}}(apis)$}
\STATE \textcolor{gray}{$archive \gets \text{\textsc{UpdateCoverage}}(\emptyset \cup testCases, covPts)$}
\WHILE{$timeElapsed < T$}
    \STATE \textcolor{gray}{$chosenTarget \gets \text{\textsc{ChooseTargetByUncover}}(archive)$}
    \STATE \textcolor{gray}{$chosenTestCase \gets \text{\textsc{ChooseTestCases}}(testCases, chosenTarget, archive)$}
    \STATE $relatedCode \gets null$
    \IF{$\text{\textsc{isBranch}}(chosenTarget) \textbf{ or } \text{\textsc{isMethodReplacement}}(chosenTarget)$ }
        \STATE $relatedCode \gets \text{\textsc{GetRelatedCode}}(chosenTarget)$
    \ENDIF
    \STATE \textcolor{gray}{$upToNTimes \gets \text{\textsc{GetNumberOfMutations}}()$}
    \STATE $llmTimes \gets 0$
    \IF{$relatedCode \neq null$}
        \STATE $llmTimes \gets \text{\textsc{Max}}(\frac{upToNTimes}{2}, M)$
    \ENDIF
    \STATE $totalTimes \gets \text{\textsc{Max}}(upToNTimes, llmTimes)$
    \FOR{$i \gets 1$ \textbf{ to } $totalTimes$}
        \IF{$i \leq llmTimes$}
            \STATE $newTestCase \gets \text{\textsc{LLMAssistedMutate}}(chosenTestCase, apis,\newline relatedCode, model)$
        \ELSE
            \STATE \textcolor{gray}{$newTestCase \gets \text{\textsc{Mutate}}(chosenTestCase, apis)$}
        \ENDIF
        \textcolor{gray}{
        \STATE $evaluateResult \gets \text{\textsc{calculateCoverage}}(archive, chosenTestCase,\newline newTestCase)$
        \IF{$\text{\textsc{CoverNotFewer}}(evaluateResult)$}
            \STATE $archive \gets \text{\textsc{UpdateCoverage}}(archive \cup newTestCase)$
            \STATE $chosenTestCase \gets \text{\textsc{UpdateCurrentTestCases}}(newTestCase)$
            \STATE $testCases \gets \text{\textsc{UpdateTestCases}}(testCases \cup newTestCase)$
        \ENDIF}
    \ENDFOR
\ENDWHILE
\STATE \RETURN $archive$
\end{algorithmic}
\end{algorithm}

\subsection{High-Level Walkthrough}
Algorithm~\ref{alg:miohint} shows \textsc{MioHint}’s high-level algorithm. \textsc{MioHint} is an extension of the MIO (Many Independent Objectives) algorithm. It incorporates an LLM to improve the mutation process for generating API test cases, with the aim of maximizing code coverage. For clarity, the parts of the algorithm that remain unchanged from the original MIO are shown in grey.

We designed \textsc{MioHint} by preserving the core framework of the MIO algorithm, including its target selection, test case evaluation, and archiving mechanisms, and integrating our LLM-assisted mutation module. This synergistic design is motivated by the crucial balance between efficiency and effectiveness. The fast random mutation is highly efficient for exploring the general test space and covering readily reachable targets. The more resource-intensive LLM is then strategically deployed as a specialist, invoked to overcome hard-to-cover targets selected by MIO that cause the search process to stagnate. The test case generated by the LLM is then reintegrated into the main search loop: it is evaluated by the MIO fitness metric coverage, and if deemed high-quality, it is archived to serve as a new, promising seed for subsequent random mutations, which further expand the exploration. This principled strategy maximizes overall testing performance by combining the broad, efficient exploration of the search algorithm with the deep, targeted problem-solving of the LLM.

The algorithm begins by extracting the set of APIs from the program under test using the function \texttt{GetApis(program)} (\textit{Line 1}). Then, it samples random test cases according to the API definition (\textit{Line 2}), calculates its coverage, and updates the coverage to the archive (\textit{Line 3}) as in Algorithm MIO. After the target is chosen, \textsc{MioHint} extracts code related to the chosen target when the target is of type branch or method replacement (\textit{Line 8-9}). We will further elaborate on how we perform our related code extraction in the next section as a vital part of our algorithm. If \texttt{relatedCode} exists, \texttt{llmTimes} is set to the maximum of half the number of mutations and a minimum count 
$M$ (\textit{Lines 14-15}). We set M to 2 in our experiment because of the randomness of responses from the LLM. We also adapt the total mutation times to LLM-assisted mutation times to ensure the LLM-assisted mutation performs at least twice (\textit{Lines 16}). We replace half of the original mutation times of vanilla mutation with our proposed LLM-assisted mutation (\textit{Lines 19}). The subsequent evaluation steps of the algorithm align with Algorithm MIO. If the new test case covers not fewer targets than the current one, update the current test case (\texttt{chosenTestCase}), as well as the \texttt{archive} and \texttt{testCases} (Lines 23-28).

\subsection{LLM-assisted Mutation}

\begin{figure}
    \centering
    \includegraphics[width=1\columnwidth]{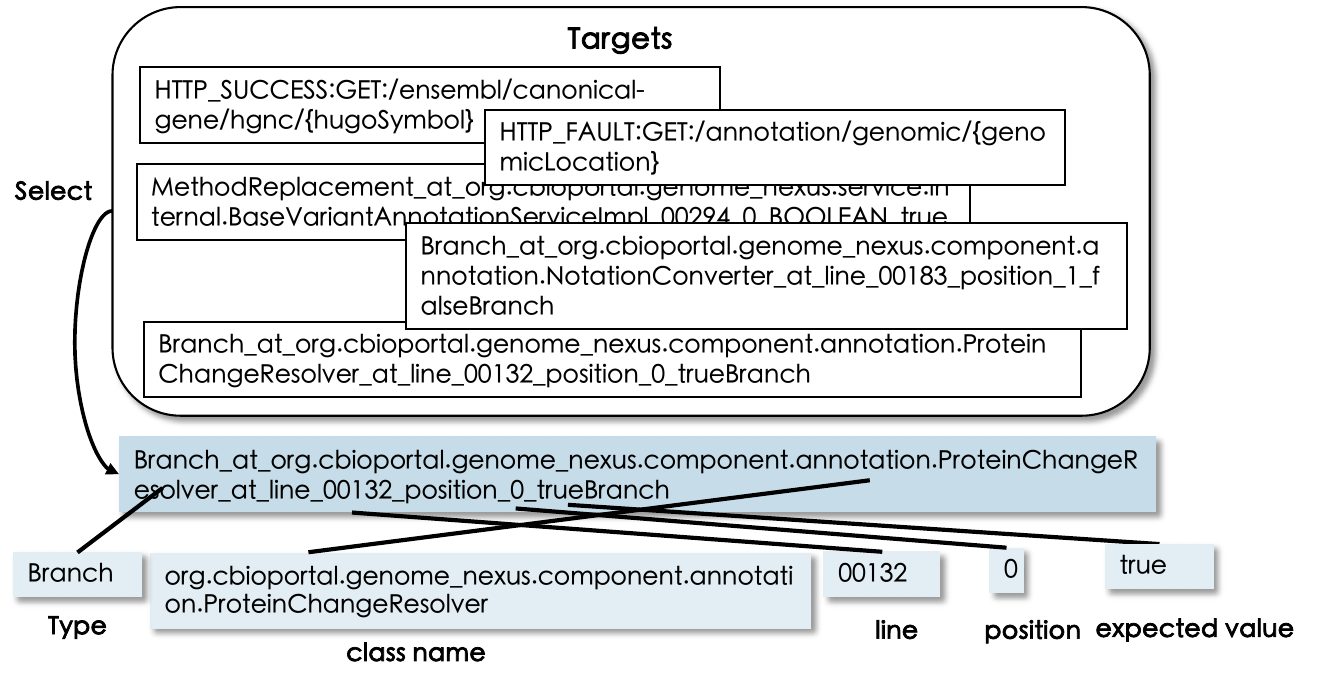}
    \caption{Targets in \textsc{MioHint}}
    \label{fig:targets}
\end{figure}

\noindent \textbf{Goal.} Given an uncovered target, retrieve relevant code information and generate a mutant with the retrieved information to cover it.

\noindent \textbf{Scope of Targets under Study.} We adhered to the target definition established in \textsc{EvoMaster} and selected a subset of targets that point to a specific line of the code for our analysis. As illustrated in Figure \ref{fig:targets}, targets encompass both branches or methods identifiable at the line code level, as well as HTTP success or failure responses applicable to the entire API. Our analysis focuses on the former category, as our objective is to maximize bytecode line coverage, while the latter can be addressed through decomposition to specific lines. 
Figure \ref{fig:targets} also illustrates the structure of targets of the type branch. Each target is defined by several components: type, class name, line number, position, and expected value. Subsequently, we extract the related code based on the components contained within these targets.

\noindent \textbf{Definition of Hard-to-cover Targets.} We define hard-to-cover targets as those selected during the MIO algorithm search process that fall within the scope of our analysis. Note that we did not perform additional filtering for these targets. These targets are characterized by the inefficiency of naive random mutation in generating the correct mutants needed to cover them, \ie numerous iterations of mutation often fail to produce a successful mutant.

\subsubsection{Related Code Extraction}
To generate a mutant that covers the target, we need to retrieve target-related code for LLM to analyze.
This part refers to Line 9 in Algorithm \ref{alg:miohint}.
The extraction is primarily divided into two parts. 
The first part is extracting the local context. It involves simply locating the specific branch or method within the target line of code, followed by identifying the function to which this line belongs. To utilize the semantic comprehension of LLM, the function includes code and annotation.
The second part is extracting the global context by value expansion. Value expansion consists of two components, def-use analysis and called function definition. 
Def-use analysis is a data flow analysis technique, and its analysis scope spans across files and methods. It begins with the variables utilized in the target, tracing back to their assignment or declaration statements. This process iteratively seeks the assignment or declaration statements for any new variables found on the right-hand side of these statements, continuing until we reach the input (request) variable. The purpose of this analysis is to establish the expression relationship between the target and the request. 
Called function definition expands the definitions of the functions called within the target line. This step is crucial because when we specify the expected return value of the method at the target line, we often need to clarify the specific content of the method to know how to produce this return value.

\begin{figure}
    \centering    
    \includegraphics[width=\columnwidth]{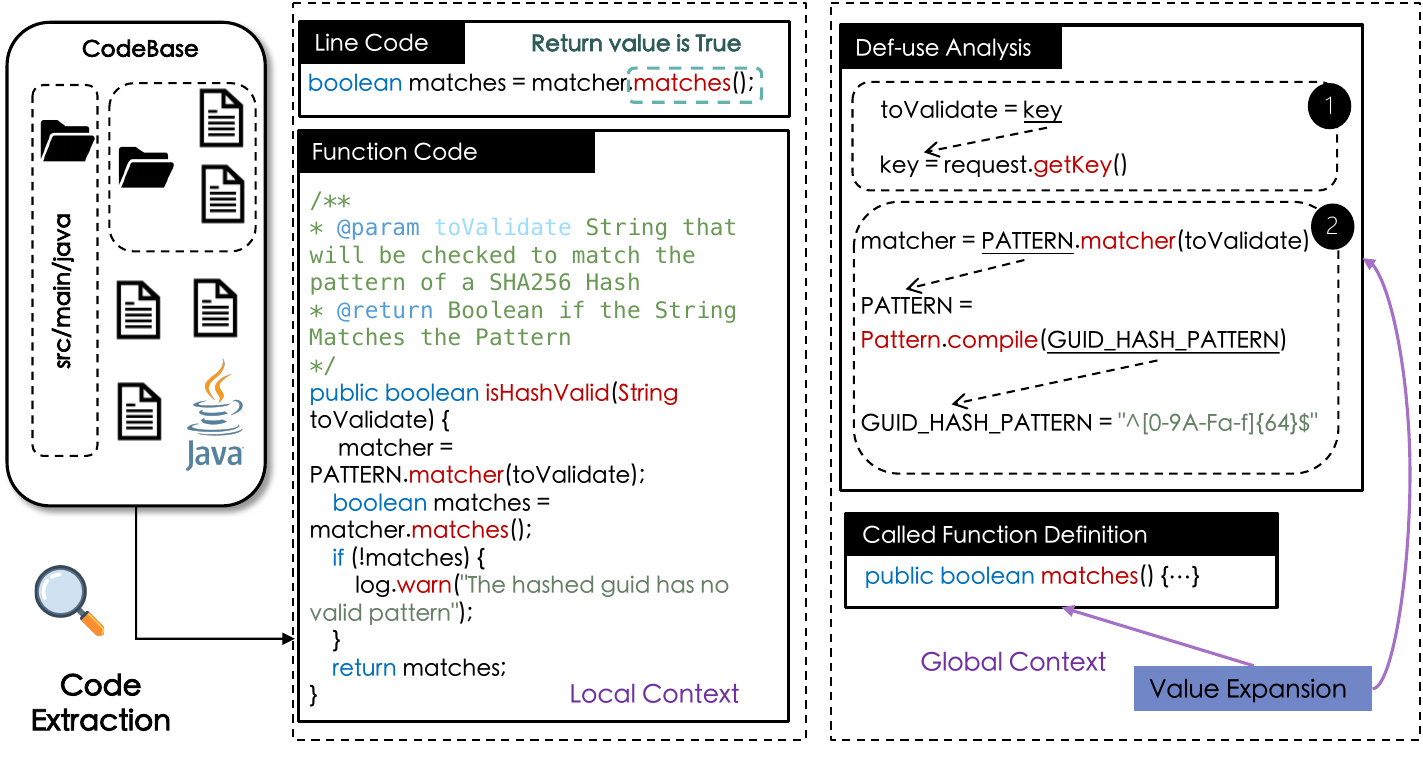}
    \caption{Code Extraction in \textsc{MioHint}}
    \label{fig:codeextraction}
\end{figure}

Figure \ref{fig:codeextraction} provides an example of the code we extract. Initially, we process the specific file containing the target line \texttt{boolean matches = matcher.matches();} and the encompassing function \texttt{public boolean isHashValid(String toValidate) \{...\}} to form the local context. 
Notably, for function code, we retain the annotations of the function to leverage the LLM's understanding of natural language. 
For global context, we perform cross-file analysis within the entire repository to get the value expansion including def-use analysis and called function expansion. 
The def-use analysis consisted of two parts. 
Part 1 of the def-use analysis is conducted within the functions at the call chain of the focal function, aiming to identify the value of the passing argument \texttt{toValidate}. When calling function \texttt{isHashValid(key)}, parameter \texttt{key} is passed, thus \texttt{toValidate = key}, and the definition of the parameter \texttt{key} is \texttt{key = request.getKey()}. 
In Part 2 of the def-use analysis, the process iteratively seeks the definition of variables used in the target. First, find the definition of the function variable \texttt{matcher} which is used at the focal line. Then find the definition of the class variable \texttt{PATTERN} imported by the definition of \texttt{matcher}. Finally, expand the definition of the class variable \texttt{GUID\_HASH\_PATTERN} imported by the definition of \texttt{PATTERN}. This iterative approach allows for a comprehensive understanding of the data flow and dependencies associated with the variables involved.

\begin{algorithm}
\footnotesize
\caption{Def-Use Analysis}
\label{alg:defuse}
\begin{algorithmic}[0]
\REQUIRE $targetStatement$
\ENSURE a chain of definition statements $defUseChain$
\STATE $defUseChain \gets \{\}$
\STATE
\STATE \textbf{function} \textsc{analyzeVariable}($variable$)
\STATE \quad $defStmt \gets \textsc{findDefinitionStatement}(variable)$
\STATE \quad \textbf{if} $defStmt$ \textbf{not in} $defUseChain$ \textbf{then}
\STATE \quad \quad $defUseChain \gets defUseChain \cup \{defStmt\}$
\STATE \quad \quad $newVariables \gets \textsc{extractVariables}(defStmt)$
\STATE \quad \quad \textbf{for each} $nv$ \textbf{in} $newVariables$ \textbf{do}
\STATE \quad \quad \quad \textsc{analyzeVariable}($nv$)
\STATE \quad \quad \textbf{end for}
\STATE \quad \textbf{end if}
\STATE \textbf{end function}
\STATE
\STATE \textbf{function} \textsc{analyzeCallers}($function$)
\STATE \quad $callers \gets \textsc{findCallers}(function)$
\STATE \quad \textbf{for each} $c$ \textbf{in} $callers$ \textbf{do}
\STATE \quad \quad $parameters \gets \textsc{extractCallParameters}(c)$
\STATE \quad \quad \textbf{for each} $p$ \textbf{in} $parameters$ \textbf{do}
\STATE \quad \quad \quad \textsc{analyzeVariable}($p$)
\STATE \quad \quad \textbf{end for}
\STATE \quad \quad \textsc{analyzeCallers}($c$)
\STATE \quad \textbf{end for}
\STATE \textbf{end function}
\STATE
\STATE $targetFunction \gets \textsc{findEnclosingFunction}(targetStatement)$
\STATE \textsc{analyzeCallers}($targetFunction$)
\STATE
\STATE $variables \gets \textsc{extractVariables}(targetStatement)$
\STATE \textbf{for each} $v$ \textbf{in} $variables$ \textbf{do}
\STATE \quad \textsc{analyzeVariable}($v$)
\STATE \textbf{end for}
\STATE \RETURN $defUseChain$
\end{algorithmic}
\end{algorithm}

We present details of def-use analysis in Algorithm \ref{alg:defuse}. 
This algorithm constructs a chain of definition statements of used variables for a given target statement according to the def-use relationship. The main procedure initializes an empty set $defUseChain$ and extracts variables from the $targetStatement$. For each variable, it calls the $analyzeVariable$ function, which recursively finds the definition statements of the variables and adds them to the $defUseChain$. The main procedure also identifies the enclosing function of the $targetStatement$ and calls the $analyzeCallers$ function with it. The $analyzeCallers$ function identifies all callers of the given function, extracts parameters of the call statement, and recursively analyzes these parameters to extend the $defUseChain$. The $analyzeCallers$ function analyzes the caller's caller iteratively until it includes all functions in the call chain. The algorithm ensures that all relevant definition statements are included in the $defUseChain$ by the end of its execution.

\indent \textbf{Analysis Scope.} The analysis is narrowly focused, helping minimize overhead. For each hard-to-cover target, the analysis scope of \textit{local context} is restricted to the file containing the target. The \textit{global context} extends this to include files traced along the call chain of the target's enclosing function.

\subsubsection{Prompt Construction}
\begin{figure}
    \centering
    \includegraphics[width=1\columnwidth]{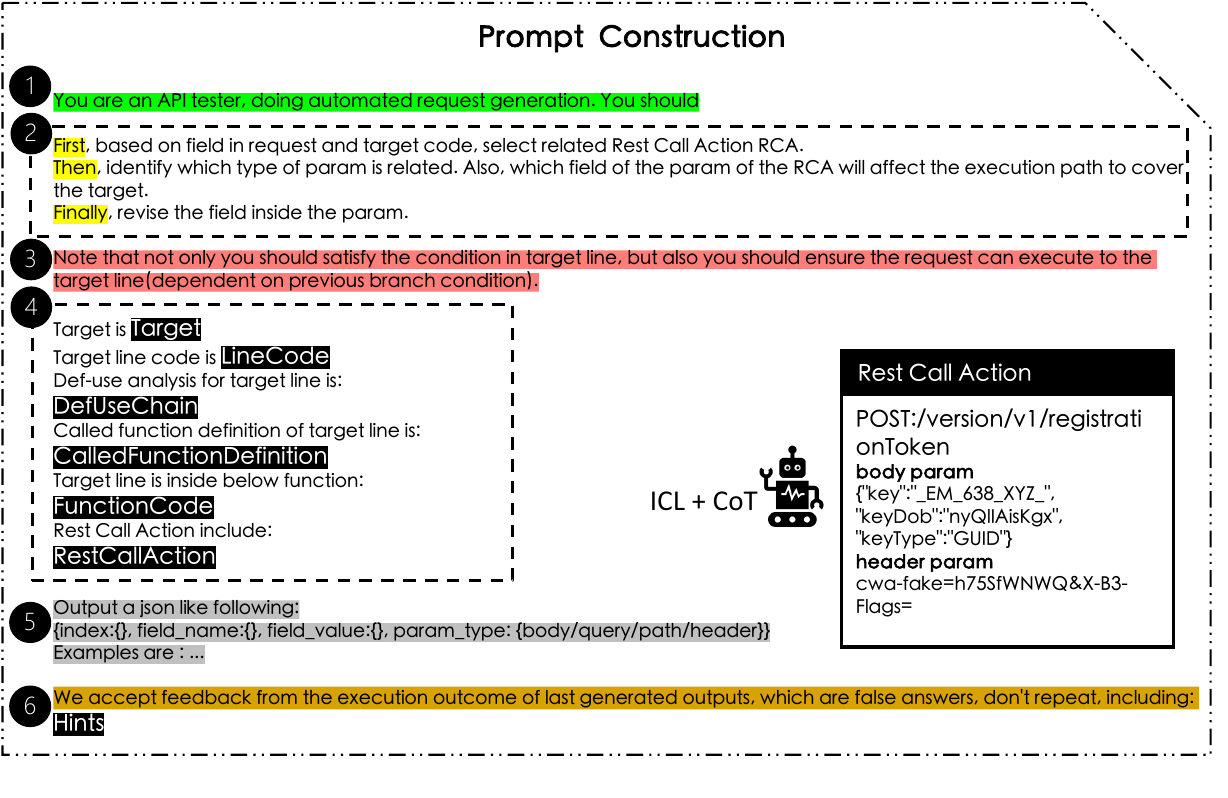}
    \caption{Prompt Construction in \textsc{MioHint}. The prompt specifies task (1), decomposes task (2), finds dependent condition (3), fills with target-specific information (4), specifies output format (5), and receives execution feedback (6).}\label{fig:promptconstruction}
\end{figure}

Having retrieved the target along with related code and REST call actions to be mutated, we construct a prompt following a structured approach.
There are six input features in the prompt, each providing orthogonal information.
As illustrated in Figure~\ref{fig:promptconstruction}, we begin with a task-specific instruction, labeled (1). The first feature defines the task, enabling LLM to understand what is required to do.
The second feature adopts a chain-of-thought (CoT) approach to stepwise locate the object to be mutated, from the request down to the specific parameter.
Specifically, this CoT involves selecting the most relevant REST call action, locating the specific type of parameter along with its field, and finally, revising the selected field, labeled (2).
To further refine the mutation process, the third feature increases attention to control flow conditions to address implicit flows, labeled (3).
Subsequently, we populate the template with previously extracted information, including the target, line of code, def-use chain, called function definition, function code, and REST call action, labeled (4).
Specifically, in the fourth feature, the Target is specified as a string (as shown in Figure~\ref{fig:targets}), indicating the target’s type, class, line, position, and expected value. LineCode is used to locate the target within the FunctionCode; the DefUseChain and CalledFunctionDefinition provide data flow relationships between the target and request; and RestCallActions represent the detailed content of the requests that can be selected for mutation.
The fifth feature defines the output format as well-defined JSON by providing illustrative examples for consistent output parsing, labeled (5).
Finally, the sixth feature incorporates hints generated from previous runs along with their execution feedback to guide the mutation process. Specifically, it leverages in-context learning to learn from historical feedback, thereby avoiding repeatedly generating incorrect mutations, labeled (6).
The absence of any of these steps can introduce errors in LLM's reasoning process.

\subsubsection{Test Case Generation}
We query the LLM using the prompt constructed in the previous section.
Its response is then post-processed to extract a mutation hint, which is represented as a JSON string.
This hint specifies the REST call action to revise, the parameter type to modify, the specific field to update, and the new value for that field.
Subsequently, the hint is applied to the original test case to generate a mutated version.
The resulting mutated test case, designed to cover the selected target, is then evaluated.
If the evaluation shows no coverage degradation, it is added to the population and becomes eligible for further mutation.
Furthermore, the evaluation result is preserved and used as feedback for subsequent LLM-assisted mutation attempts on the same target.

\section{Evaluation}\label{sec:evaluation}
In this section, we evaluate our method using real-world programs and answer the following questions:
\begin{itemize}[leftmargin=*]
\item  \textbf{RQ1:} How effective is \textsc{MioHint} in generating test cases to cover the SUT?
\item  \textbf{RQ2:} To what extent does the value expansion component improve the test generation effectiveness of \textsc{MioHint}?
\item  \textbf{RQ3:} What is the runtime efficiency of \textsc{MioHint}?

\end{itemize}

\begin{table}[]
    \vspace{0.15in}
    \footnotesize
    \centering
    \caption{\small Real-world benchmark programs used in the evaluation.}
    \label{tab:bm-info}
    \begin{tabular}{l|r|r|r}
        \toprule[2pt]
        \textbf{SUT} & \textbf{\#SourceFiles} & \textbf{\#LOCs}    & \textbf{\#Endpoints}        \\
        \midrule[1pt]
        \textit{catwatch}               & 106              &  9636   & 14 \\
        \textit{cwa-verification}      & 47                  & 3955        & 5  \\
        \textit{features-service} & 39 & 2275        & 18 \\
        \textit{genome-nexus} & 405 & 30004        & 23 \\
        \textit{gestaohospital} & 33 & 3506        & 20 \\
        \textit{languagetool} & 1385 & 174781        & 2 \\
        \textit{market} & 124 & 9861        & 13 \\
        \textit{ocvn} & 526 & 45521        & 258 \\
        \textit{proxyprint} & 73 & 8338        & 74 \\
        \textit{rest-ncs} & 9 & 605        & 6 \\
        \textit{rest-scs} & 13 & 862        & 11 \\
        \textit{restcountries} & 24 & 1977        & 22 \\
        \textit{scout-api} & 93 & 9736        & 49 \\
        \textit{pay-publicapi} & 232 & 12044        & 10 \\
        \textit{reservations-api} & 31 & 846        & 7 \\
        \textit{session-service} & 14 & 468        & 8 \\
        \bottomrule[2pt]
    \end{tabular}
\end{table}

\subsection{Evaluation Setup}
\noindent\textbf{\textit{BaseLine.}}
We compare our method with a state-of-the-art whitebox API testing tool Evo\textsc{Master} with the latest version 3.2.0, which is publicly available by the time of writing this paper.

\noindent\textbf{\textit{Evaluation Metrics.}}
We use five metrics to evaluate the performance of different configurations. All statistics presented are average values derived from repeated experiments.
\begin{itemize}[leftmargin=*]
\item \textit{Line coverage}, which refers to bytecode line coverage of the repository, measures the extent to which the bytecode generated from source code is executed during testing.
\item \textit{Target coverage}, which refers to the coverage of selected hard-to-cover targets. Only branch or method target is in this scope.
\item \textit{Mutation hit rate}, which refers to the proportion of mutations that successfully cover the target.
\item \textit{Number of mutation times}, this metric represents the total number of mutations performed during the experiment.
\item \textit{Average execution time per test}, this metric represents the execution time of every run of the test case mutation and evaluation.
\end{itemize}

\noindent\textbf{\textit{Evaluation Datasets.}}
We use real-world programs from the EMB~\cite{emb} corpus as the evaluation benchmarks. EMB is a well-maintained open-source corpus of Web APIs (including REST, GraphQL, and RPC APIs). The detail of the real-world benchmark programs we used is shown in Table~\ref{tab:bm-info}, including the number of source files, lines of code, and the number of REST endpoints in each API. These statistics only account for the business logic code and exclude third-party libraries (e.g., HTTP servers).

EMB offers APIs of varying sizes and complexities from different domains, covering a diverse set of APIs needed for scientific experimentation. There are two artificial APIs designed to study numeric (\textit{rest-ncs}) and string (\textit{rest-scs}) constraints. The other 14 APIs are sourced from GitHub: some are from public administrations (e.g., \textit{ocvn}), while others are popular tools providing a REST interface (e.g., \textit{languagetool}). 

Some SUTs were excluded due to objective factors. Specifically, the \textit{features-service} encountered a runtime error. Additionally, all targets within the \textit{cwa-verification}, \textit{scout-api}, and \textit{session-service} are related to database operations or configurations, which modify requests that can not be addressed. Furthermore, the \textit{ocvn}, \textit{proxyprint}, \textit{market}, and \textit{reservations-api} do not contain any targets within the scope of analysis during the search process.

\noindent\textbf{\textit{Experiment Settings.}}
Each fuzzing session runs for 1 hour. To account for the randomness of search-based fuzzing, each experiment was repeated 10 times. With 3 settings and 16 SUTs, it required 3 × 16 × 10 = 480 hours, \ie 20 days of computation.
We conducted our evaluations on the machine equipped with an Intel Xeon Gold 5218R CPU with 10 cores, using Ubuntu 20.04.6 LTS as the operating system. 
For LLM query, we use public APIs provided by OpenAI with GPT-4o~\cite{openai}.

\begin{table*}[!ht]
\small
\centering
\caption{Performance comparisons between the Baseline and Our Method, in terms of average (i.e., arithmetic mean) line coverage, num of target, target coverage, and mutation hit rate. Results of statistical tests are reported, including $p$-values using Mann-Whitney-Wilcoxon U-tests and $\hat{A}_{12}$ effect sizes using Vargha-Delaney statistics. For p-values lower than the threshold $\alpha = 0.05$, the effect sizes $\hat{A}_{12}$ are shown in bold. Base = baseline, MH = \textsc{MioHint}.}
\begin{tabular}{lccccccccccc}
\toprule
\multirow{2}{*}{\textbf{SUT}} & \multicolumn{4}{c}{\textbf{Line Coverage \%}}  & \multicolumn{2}{c}{\# \textbf{Num of Target}} & \multicolumn{2}{c}{\textbf{Target Coverage \%}} & \multicolumn{2}{c}{\textbf{Mutation Hit Rate \%}} \\
\cmidrule(lr){2-5} \cmidrule(lr){6-7} \cmidrule(lr){8-9} \cmidrule(lr){10-11}
 & Base & MH & $\hat{A}_{12}$ & p-value & Base & MH & Base & MH & Base & MH \\
\midrule
\textit{catwatch} & 42.50 & 45.40 & \textbf{0.05} & <0.001 & 22 & 22 & 0.00 & 29.40 & 0.00 & 5.20 \\
\textit{genome-nexus} & 34.60 & 39.00 & \textbf{0} & <0.001 & 31 & 31 & 4.56 & 54.74 & 0.18 & 12.23 \\
\textit{gestaohospital} & 45.10 & 45.20 & 0.45 & 0.582 & 5 & 6 & 4.76 & 46.82 & 0.17 & 9.14 \\
\textit{languagetool} & 23.90 & 35.00 & \textbf{0.01} & <0.001 & 118 & 81 & 0.22 & 23.48 & 0.02 & 7.80 \\
\textit{pay-publicapi} & 13.00 & 12.90 & 0.55 & 0.368 & 5 & 2 & 0.00 & 100.00 & 0.00 & 100.00 \\
\textit{rest-ncs} & 90.30 & 92.00 & 0.18 & 0.014 & 25 & 29 & 20.08 & 40.47 & 0.77 & 4.44 \\
\textit{rest-scs} & 69.80 & 87.80 & \textbf{0} & <0.001 & 48 & 92 & 14.97 & 92.30 & 0.42 & 28.16 \\
\textit{restcountries} & 68.90 & 70.40 & 0.2 & 0.016 & 47 & 35 & 33.98 & 73.07 & 1.27 & 10.43 \\
\midrule
\textbf{Total} & 48.51 & 53.46 & 0.18 & &  & 31 & 9.82 & 57.54 & 0.35 & 22.17 \\
\bottomrule
\end{tabular}
\label{tab:rq1}
\end{table*}

\subsection{RQ1: Effectiveness on Covering SUT}
To assess the effectiveness of our LLM-assisted design, we conduct a comparative evaluation against a baseline across three key metrics: \textit{line coverage}, \textit{target coverage}, and \textit{mutation hit rate}. Specifically, \textit{line coverage} serves as a general indicator of the overall code coverage achieved by both our method and the baseline. Higher coverage values indicate that \textsc{MioHint} is more effective at generating test cases that thoroughly exercise the SUTs. Furthermore, \textit{target coverage} specifically quantifies our method's effectiveness in covering hard-to-cover targets. Lastly, \textit{mutation hit rate} assesses the accuracy and effectiveness of our LLM-assisted mutation strategy compared to naive mutation. Table~\ref{tab:rq1} shows the results in detail for our method compared to the baseline. We follow the statistical guidelines from ~\cite{10.1145/1985793.1985795}, reporting $p$-values of Mann-Whitney-Wilcoxon U tests and Vargha-Delaney standarized $\hat{A}_{12}$ effect sizes.

For metric \textit{line coverage}, our method improves line coverage for most SUTs. Results are statistically significant for 4 SUTs, with no statistically worse results. On these APIs improvement are either "medium" (\eg +2.9\% for \textit{catwatch} and +4.4\% for \textit{genome-nexus}) or "large" (\eg +11.1\% for \textit{languagetool} and +18.0\% for \textit{rest-scs}). Overall, our method achieved an average increase in line coverage of 4.95\% (from 48.51\% to 53.46\%). This indicates that our method contributes not only to the selected target but also to the overall performance of the search algorithm.

In terms of \textit{target coverage}, our method demonstrates even more substantial improvements. For instance, the target coverage for \textit{genome-nexus} increased dramatically from 4.56\% to 54.74\%, and for \textit{pay-publicapi}, it soared from 0.00\% to 100.00\%. On average, our method achieved an increase in target coverage of 47.72 percentage points (from 9.82\% to 57.54\%). This indicates that our LLM-assisted mutation can cover more than half of the hard-to-cover targets.

To clarify the relationship between line coverage and target coverage, and to explain anomalies such as the case of the service \textit{pay-publicapi}, where target coverage increases from 0\% to 100\% without a corresponding rise in line coverage, it is important to consider the underlying factors. Line coverage is closely related to the total number of covered targets. Our mutation strategy is designed to maximize the coverage of selected targets, but when the number of targets is low, even full target coverage may not significantly increase overall line coverage. For example, \textit{pay-publicapi} has only two selected targets. This is due to the target selection strategy determined by the underlying MIO search-based algorithm. As a result, target coverage can increase substantially while line coverage remains largely unchanged. Similarly, for the service \textit{gestaohospital}, the nearly identical line coverage under both settings is also due to the small number of targets, which is five or six.

\textit{Mutation hit rate} is a crucial metric for assessing the quality of test cases generated by mutation. Our method also shows significant improvements in this metric. For example, the mutation hit rate for \textit{genome-nexus} increased from 0.18\% to 12.23\%, and for \textit{rest-scs}, it rose from 0.42\% to 28.16\%. Overall, our method achieved an average increase in mutation hit rate of 21.82 percentage points (from 0.35\% to 22.17\%). This indicates that the accuracy of our LLM-assisted mutation is \textbf{67$\times$} higher than that of the vanilla mutation. 

For our studied targets, which are identified as hard-to-cover due to their low coverage and mutation hit rates in the baseline, typically lower than 5\% and 0.2\%, respectively, for most of the systems under test (SUTs), our method shows significant improvement in these two metrics, to 57.54\% and 22.17\% on average.

\vspace{5pt}
\noindent
\colorbox{gray!10}{
    \parbox{\dimexpr\linewidth-2\fboxsep-2\fboxrule\relax}{
        \textbf{Answer to RQ1: } Our method consistently outperforms the baseline across all metrics: average line coverage (53.46\% vs. 48.51\%), an increase of 4.95\%; target coverage (57.54\% vs. 9.82\%), more than half of the targets; and mutation hit rate (67$\times$).
    }
}

\begin{table}[t]
\small
\centering
\caption{Comparison of Baseline, Our Method, and no VE in terms of average line coverage (LC), target coverage (TC), and mutation hit rate (MHR).}
\begin{tabular}{l p{0.3cm} p{0.3cm} p{0.5cm}  p{0.3cm} p{0.3cm} p{0.5cm}  p{0.3cm} p{0.3cm} p{0.5cm}}
\toprule
\multirow{2}{*}{\textbf{SUT}} & \multicolumn{3}{c}{\textbf{Baseline}} & \multicolumn{3}{c}{\textbf{Our Method}} & \multicolumn{3}{c}{\textbf{no VE}} \\
\cmidrule(lr){2-4} \cmidrule(lr){5-7} \cmidrule(lr){8-10}
 & \textbf{LC} & \textbf{TC} & \textbf{MHR} & \textbf{LC} & \textbf{TC} & \textbf{MHR} & \textbf{LC} & \textbf{TC} & \textbf{MHR} \\
\midrule
\textit{catwatch} & 42.5 & 0.0 & 0.0 & 45.4 & 29.4 & 5.2 & 44.7 & 27.2 & 4.7 \\
\textit{genome-nexus} & 34.6 & 4.6 & 0.2 & 39.0 & 54.7 & 12.2 & 38.0 & 54.5 & 15.1 \\
\textit{gestaohospital} & 45.1 & 4.8 & 0.2 & 45.2 & 46.8 & 9.1 & 45.0 & 42.2 & 10.4 \\
\textit{languagetool} & 23.9 & 0.2 & 0.0 & 35.0 & 23.5 & 7.8 & 28.7 & 24.6 & 7.8 \\
\textit{pay-publicapi} & 13.0 & 0.0 & 0.0 & 12.9 & 100.0 & 100.0 & 13.0 & 50.0 & 27.8 \\
\textit{rest-ncs} & 90.3 & 20.1 & 0.8 & 92.0 & 40.5 & 4.4 & 89.3 & 35.3 & 4.2 \\
\textit{rest-scs} & 69.8 & 15.0 & 0.4 & 87.8 & 92.3 & 28.2 & 87.7 & 95.1 & 25.8 \\
\textit{restcountries} & 68.9 & 34.0 & 1.3 & 70.4 & 73.1 & 10.4 & 70.0 & 62.8 & 7.4 \\
\midrule
\textbf{Total} & 48.5 & 9.8 & 0.4 & 53.5 & 57.5 & 22.2 & 52.0 & 49.0 & 12.9 \\
\bottomrule
\end{tabular}
\label{tab:rq2}
\end{table}

\subsection{RQ2: Effectiveness of Value Expansion}
To demonstrate the effectiveness of our value expansion technique, we conduct an ablation study on value expansion and evaluate it using the same metrics in RQ1. Table \ref{tab:rq2} shows the results in detail for the performance between three different configurations: the baseline, our proposed method, and our method without variable expansion (no VE).

After disabling value expansion, line coverage decreases by 1.5\%, target coverage decreases by 8.5\%, and the mutation hit rate drops from 22.2\% to 12.9\%, which is approximately half of the original rate. The decrease in performance across all metrics highlights the effectiveness of our value expansion. This indicates that with our value expansion, an additional 8.5\% of the hard-to-cover targets can be addressed; these targets are more challenging than the 48.5\% targets covered without value expansion, as they need more contextual information to build helpful mutation hints.

\vspace{5pt}
\noindent
\colorbox{gray!10}{
    \parbox{\dimexpr\linewidth-2\fboxsep-2\fboxrule\relax}{
        \textbf{Answer to RQ2:} Our value expansion approach leads to a 1.5\% absolute improvement in line coverage and enables us to cover 8.5\% more challenging targets that require contextual mutation hints.
    }
}

\subsection{RQ3: Runtime Efficiency}
To evaluate the runtime overhead introduced by LLM-assisted mutation, we use \textit{Number of Mutation Times} and \textit{Average Execution Time per Test} as metrics. Since the duration of each test is fixed at 1 hour, this metric can measure the number of mutations performed within the same time under different experimental settings. By comparing the number of mutations, we can assess the runtime overhead: the more mutations performed within the same time, the lower the time cost per mutation, indicating a smaller runtime overhead. We also calculate \textit{Average Execution Time per Test} to directly measure the time cost of every run of test case mutation and evaluation. Table \ref{tab:rq3} shows the results in detail for the runtime overhead between three different configurations: the baseline, our proposed method, and our method without variable expansion (no VE).

It shows that our method causes slightly more runtime overhead than the baseline. Our method got an 18\% reduction of mutation times from the baseline. In terms of average execution time per test, our method generally results in longer execution times compared to the baseline, which aligns with the intuition that the query of a language model (LLM) is expected to incur additional runtime overhead. However, the increase in runtime overhead is relatively modest, amounting to only 31\% of the baseline.

In summary, our method introduces additional runtime overhead.
However, this trade-off is justified by the considerable improvement in accuracy.
As a result, although the number of mutations is reduced, the higher accuracy of each mutation leads to an overall increase in line coverage.

Next, we discuss the runtime overhead introduced by value expansion. Interestingly, we found that disabling value expansion actually increased the overhead (from 131\% to 140\%). This is because, without the auxiliary information provided by value expansion, the accuracy of LLM-assisted mutation decreases, leading to an increased number of LLM queries. That is because, when LLM-assisted mutation can not generate a correct mutation to cover the target, more queries will be initiated to continue the search process. Consequently, the overall overhead increases. This finding highlights the importance of value expansion, as it incurs minimal overhead while significantly enhancing accuracy. Given the considerable time cost associated with LLM queries, it becomes necessary to maximize the benefits of each query by constructing prompts with more accurate information.

\vspace{5pt}
\noindent
\colorbox{gray!10}{
    \parbox{\dimexpr\linewidth-2\fboxsep-2\fboxrule\relax}{
        \textbf{Answer to RQ3: } Our method increases runtime overhead per test by 31\%. Value expansion enhances the accuracy of LLM-assisted mutation, further reducing the LLM query times and contributing to the decrease in overall runtime overhead.
    }
}

\begin{table}[t]
\small
\centering
\caption{Comparison of Mutation Times and Execution Time per Test between different configurations: Base = baseline, All = our method, no VE = our method without value expansion}
\begin{tabularx}{\linewidth}{lcccccc}
\toprule
\multirow{2}{*}{\textbf{SUT}} & \multicolumn{3}{c}{\textbf{\# Mutations}} & \multicolumn{3}{c}{\textbf{Avg. Time per Test (ms)}} \\
\cmidrule(lr){2-4} \cmidrule(lr){5-7}
 & \textbf{Base} & \textbf{All} & \textbf{no VE} & \textbf{Base} & \textbf{All} & \textbf{no VE} \\
\midrule
\textit{catwatch} & 1336 & 1115 & 1038 & 2918 & 3513 & 3848 \\
\textit{genome-nexus} & 1531 & 1111 & 1202 & 2523 & 3583 & 3314 \\
\textit{gestaohospital} & 2183 & 1940 & 1893 & 1711 & 2008 & 1983 \\
\textit{languagetool} & 2034 & 1073 & 1024 & 1866 & 3793 & 4574 \\
\textit{pay-publicapi} & 2394 & 2361 & 2371 & 1536 & 1559 & 1550 \\
\textit{rest-ncs} & 2505 & 2261 & 2067 & 1520 & 1630 & 1831 \\
\textit{rest-scs} & 2501 & 2095 & 1740 & 1527 & 1845 & 2232 \\
\textit{restcountries} & 2498 & 1969 & 1964 & 1527 & 1944 & 1898 \\
\midrule
\multirow{2}{*}{\textbf{Total}}  & 2123 & 1741 & 1662 & 1891 & 2484 & 2654 \\
 & (100\%) & (82\%) & (78\%) & (100\%) & (131\%) & (140\%) \\
\bottomrule
\end{tabularx}
\label{tab:rq3}
\end{table}

\section{Threats to Validity}
\noindent \textbf{Internal.}
A threat to the internal validity of our work arises from the randomness inherent in MIO and LLMs.
To mitigate this threat, we repeated each experiment 10 times.
Another threat relates to our dependency on the underlying search algorithm.
Specifically, we retained the original target selection strategy of the MIO algorithm, which may not be optimal for our LLM-assisted mutation approach.
In our experiments, this led to the exclusion of four benchmark programs because the process failed to identify any targets within the designated analysis scope.
For future work, we plan to co-design our mutation strategy with other components of the MIO algorithm to improve overall performance.

\noindent \textbf{External.}
The generalizability of our tool's testing ability might be limited by several factors. One comes from our evaluations, which only focus on open-source services written in Java and utilize specific language models, GPT-4o. The risk of generalizability across different programming languages can be resolved by the same comprehension ability of ChatGPT across different programming languages (\eg Java, JavaScript, Python, and C)~\cite{10.1145/3650212.3680399}. Besides, def-use analysis is a common static analysis that can be applied to other languages. These indicate that our method is compatible with different languages.
Another risk comes from the metric we select. Although there is a very strong correlation between the coverage achieved and the number of bugs found by a fuzzer, fuzzers best at achieving coverage, may not be best at finding bugs~\cite{10.1145/3510003.3510230}. Nevertheless, given that coverage is the most widely used metric, we continue to employ it.
Additionally, the requirement for source code access may limit the applicability of our tool in scenarios where source code is not available, thus affecting its generalizability to closed-source or proprietary software.




\section{Related Work}

\noindent \textbf{Automated API Testing.} Blackbox API testing analyzes API specifications to infer dependencies between parameters. \textsc{RESTler}~\cite{10.1109/ICSE.2019.00083} generates the correct sequence of API requests by producer-consumer dependencies. \textsc{Morest}~\cite{liu2022morestmodelbasedrestfulapi} generates a RESTful Service Property Graph (RPG) using extracted dependencies between APIs. Besides, there are other works resolving oracle problem~\cite{10.1145/3180155.3182528,10.1007/978-3-030-38991-8_10}, handling database~\cite{10.1007/978-3-030-88106-1_8}, and so on. 
Recent advances have explored using LLMs to further improve blackbox API testing, including: API specifications enhancement by identifying rules and generating example parameter values using LLM~\cite{kim2024leveraging}; automated specifications inference using LLM for black-box testing even without realistic specification~\cite{decrop2024you}; and multi-agent reinforcement learning that incorporates a semantic property dependency graph alongside LLMs to simplify the search space for dependencies and optimize API exploration~\cite{kim2024multi}.
Whitebox API testing focuses on the design of heuristics~\cite{Arcuri_2017}. To solve the common issue in SBST, the flag problem~\cite{10.1007/3-540-45110-2_148}, testability transformations~\cite{arcuri2021enhancing} are proposed to transform the SUT’s source code in a way that enhances the fitness function. Our work is orthogonal to the enhancement of fitness function, as our focus is on improving the accuracy of mutation operations. Therefore, these two aspects can complement each other in a synergistic manner.

\noindent \textbf{LLMs for Unit Test Generation.} Current efforts to enhance the effectiveness of unit test generation with LLMs encompass several strategies. These include pre-training or fine-tuning of LLMs~\cite{alagarsamy2023a3testassertionaugmentedautomatedtest,shin2024domainadaptationcodemodelbased,khanfir2023efficientmutationtestingpretrained}, the design of effective prompts using focal context~\cite{chen2024chatunitestframeworkllmbasedtest}, including focal method and focal class or continuous improvement by a generator and refiner~\cite{yuan2024manualtestsevaluatingimproving}, and test generation with additional documentation, which incorporates API documentation~\cite{vikram2024largelanguagemodelswrite} and user-written bug reports~\cite{plein2023automaticgenerationtestcases}. Furthermore, the integration of LLMs with traditional Search-Based Software Testing (SBST) is being explored~\cite{codamosa2023}. In instances where SBST stalls, the LLM can be queried for sample test cases.
These approaches are fundamentally limited by their narrow scope. Designed for unit testing, their analysis is inherently confined to isolated code snippets like individual functions. In contrast, our work targets system-level API testing, which demands a holistic understanding of interactions spanning the entire repository. Consequently, these unit-level techniques are insufficient as they overlook the broader codebase context required to understand the critical cross-module dependencies and end-to-end behaviors.

\noindent \textbf{LLMs for Repository-Level Task.}
Repository-Level tasks, including package migration~\cite{bairi2023codeplanrepositorylevelcodingusing}, issue resolution~\cite{ma2024understand,xia2024agentless,zhang2024autocoderover}, and code completion~\cite{phan2024repohyper,liu2024graphcoder}, necessitate a thorough analysis of the entire code repository. The objective of this analysis is to excavate and understand the intricate web of dependencies that exist within the repository. This broader perspective allows for a more holistic understanding of the codebase, enabling more effective and efficient solutions to the tasks at hand. 
\textsc{CodePlan} represents these dependencies as a dependency graph, encompassing syntactic relations, import relations, inheritance relations, method override relations, method invocation relations, object instantiation relations, and field use relations.
\textsc{RepoUnderstander} delves into these dependencies by constructing a repository knowledge graph. \textsc{Agentless}, on the other hand, employs the repository structure format to construct a succinct representation of the repository's file and directory structure.
\textsc{RepoHyper} constructs the Repo-level Semantic Graph (RSG), a novel semantic graph structure that encapsulates the vast global context of code repositories at the method level. This graph includes import relations, invoke relations, ownership relations, encapsulation relations, and class hierarchy relations.

While repository-level approaches recognize the importance of a global perspective, they differ in granularity and purpose. Typically, they build dependency or semantic graphs at the method or class level, resulting in coarse representations that are suitable for tasks like code completion or migration but inadequate for system-level testing. Such methods often include irrelevant code, obscuring the critical value transmission path from request to target. In contrast, our approach conducts statement-level data flow analysis, enabling the extraction of precise and minimal code slices that accurately capture the value propagation of an API call. This fine-grained context is essential for effective test generation and targeted mutation, offering a level of precision not achieved by existing techniques.

\section{Conclusion}
We present \miohint, an LLM-assisted request mutator designed to address the fitness plateau issue associated with inefficient random mutation. \miohint generates precise mutation hints using LLMs with its code comprehension ability. To identify the relationship between requests and target conditions, \miohint performs cross-file, cross-procedure data flow analysis to gather global context for the target, which enables more accurate targeted mutation guidance.
Evaluation results on 16 real-world REST API services across variant functionalities demonstrate that \miohint outperforms the state-of-the-art white-box API testing tool \textsc{EvoMaster} by 67$\times$ in mutation accuracy and achieves 4.95\% higher line coverage. Furthermore, our method is able to cover over 57\% of hard-to-cover targets, whereas the baseline achieves coverage of less than 10\%.

\bibliographystyle{ACM-Reference-Format}
\bibliography{main}

\appendix

\end{document}